\documentclass{llncs}
\usepackage[usenames,dvipsnames]{color}
\usepackage{algorithmic}
\usepackage{algorithm}
\usepackage{graphicx}
\usepackage{multirow}
\usepackage{cases}

\DeclareGraphicsExtensions{.jpg}

\newcommand{\secref}[1]{section~\ref{sec_#1}}
\newcommand{\Secref}[1]{Section~\ref{sec_#1}}

\newcommand{\figref}[1]{Fig.~\ref{fig_#1}}
\newcommand{\tblref}[1]{Table~\ref{tbl_#1}}
\newcommand{\eqref}[1]{Eq.~(\ref{eq_#1})}
\newcommand{\algref}[1]{Alg.~\ref{alg_#1}}

\newcommand{\argmax}[0]{\arg\!\max}

\begin{document}

%
%

\title{Unfolding network communities by combining defensive and offensive label propagation}
\titlerunning{Unfolding network communities}
\author{Lovro \v Subelj\thanks{Corresponding author: lovro.subelj@fri.uni-lj.si.} \and Marko Bajec}
\authorrunning{Lovro \v Subelj and Marko Bajec}
\institute{Laboratory for Data Technologies, University of Ljubljana, Ljubljana, Slovenia}

\maketitle

%
%

\begin{abstract}
Label propagation has proven to be a fast method for detecting communities in complex networks. Recent work has also improved the accuracy and stability of the basic algorithm, however, a general approach is still an open issue. We propose different label propagation algorithms that convey two unique strategies of community formation, namely, defensive preservation and offensive expansion of communities. Furthermore, the strategies are combined in an advanced label propagation algorithm that retains the advantages of both approaches; and are enhanced with hierarchical community extraction, prominent for the use on larger networks. The proposed algorithms were empirically evaluated on different benchmarks networks with planted partition and on over $30$ real-world networks of various types and sizes. The results confirm the adequacy of the propositions and give promising grounds for future analysis of (large) complex networks. Nevertheless, the main contribution of this work is in showing that different types of networks (with different topological properties) favor different strategies of community formation.
\end{abstract}

%
%

\section{\label{sec_intro}Introduction}
Complex networks commonly comprise of local structural \textit{modules} or \textit{communities} that are groups of nodes strongly connected within and only weakly connected with the rest of the network. These modules play crucial roles in many real-world systems~\cite{GD03,PDFV05}, moreover, they provide an important insight into structure and function of (large) complex networks~\cite{PDFV05,Str01,LLDM08}.

Over the last decade the research community has shown a considerable interest in detecting communities in real-world networks. Thus, a number of approaches has been presented in the literature. In particular, approaches optimizing \textit{modularity}\footnotemark[1] $Q$~\cite{CNM04,BGLL08,BC09}, graph partitioning~\cite{GN02,RCCLP04,PSSL09} and spectral~\cite{DM04,New06a} algorithms, statistical methods~\cite{NL07}, algorithms based on dynamic processes~\cite{RAK07,RB08,PSSL09,RN10}, overlapping, hierarchical and multiresolution methods~\cite{PDFV05,Gre09,RN10}, and other~\cite{LM09b,LM09c} (for a thorough review see~\cite{For10}). 

\footnotetext[1]{Significance of communities due to a selected \textit{null model}~\cite{NG04}.} 

Due to the size of large real-world networks recent research has focused on developing scalable algorithms that can be applied to networks with several millions of nodes and billions of edges. Raghavan~et~al.~\cite{RAK07} proposed using simple \textit{label propagation}, where labels are propagated among nodes until an equilibrium is reached. The main advantage of the label propagation is its near linear time complexity (in the number of edges of the network); however, due to the algorithm's simplicity, the accuracy of revealed community structure is often not state-of-the-art. 

The basic algorithm was further analyzed in~\cite{TK08} and refined into a modularity optimization algorithm in~\cite{BC09,LM09b}. Extension to directed networks was considered in~\cite{LHLC09}. Furthermore, Leung~et~al.~\cite{LHLC09} improved the basic label propagation by applying label \textit{hop attenuation} and \textit{node preference} (i.e. node propagation strength). We proceed their work in developing two unique strategies of community formation, namely, \textit{defensive preservation} of communities, where preference is applied to the core of each community, and \textit{offensive expansion} of communities, where preference is applied to the border of each community. Moreover, the two strategies are combined into an advanced label propagation algorithm (denoted \textit{K-Cores}) that preserves the advantages of both approaches. For the use with larger networks, we also present two different manners of \textit{hierarchical} community extraction.\footnotemark[2]

\footnotetext[2]{The work presented in this article was already (partially) presented in~\cite{SB00}.}

Proposed algorithms were rigorously analyzed on different benchmark networks with planted partition and on a large number of real-world networks of various types and sizes. The results justify the adequacy of the propositions and give promising grounds for future analysis of (large) complex networks. Furthermore, the analysis also shows that the appropriateness of the strategies of community formation strongly correlates with the type of the network (i.e. with its topological properties).

The rest of the article is structured as follows. \Secref{lpa} gives a formal presentation of label propagation and briefly surveys relevant subsequent refinements of the basic algorithm. Defensive and offensive strategies of community formation, and corresponding algorithms, are presented and discussed in \secref{defoff}. Empirical evaluation with discussion is done in \secref{eval} and conclusion in \secref{conc}.

\section{\label{sec_lpa}Label propagation and advances}
Let the network be represented by an undirected (multi-)graph $G(N,E)$, where $N$ is the set of nodes and $E$ is the set of edges. Furthermore, let $w_{nm}$ be the weight of the edge between nodes $n$ and $m$, $n,m\in N$. Next, denote $c_n$ to be the community (label) of node $n$ and $\mathcal{N}_n$ the set of its neighbors. Moreover, denote $\mathcal{N}_n^l$ to be the set of neighbors of $n$ that share label $l$.

\textit{Label propagation algorithm} (\textit{LPA})~\cite{RAK07} reveals network communities by employing the following procedure. At first each node $n\in N$ is labeled with an unique label, $c_n=l_n$. Next, at each iteration, each node adopts the label shared by most of its neighbors. Hence,
\begin{eqnarray}
c_n=\argmax_l|\mathcal{N}^l_n|,
\label{eq_lpa}
\end{eqnarray}
where in the case of ties one of the labels is selected at random (node $n$ retains its current label, when it is among most frequent in $\mathcal{N}_n$). The process continues until none of the labels change anymore, i.e. an equilibrium is reached. During the course of the algorithm, densely connected sets of nodes form a consensus on some particular label; thus, at the end, nodes sharing the same label are classified into the same community.

Leung~et~al.~\cite{LHLC09} have observed that basic label propagation applied to large (web) graphs commonly produces one \textit{major community} that occupies most of the nodes. However, they have shown that the emergence of a major community can be eliminated by using label \textit{hop attenuation} technique. Each label $l_n$ has associated an additional score $s_n$ (initially set to $1$) that decreases by $\delta$ after each propagation ($\delta$ is an \textit{attenuation ratio}). When $s_n$ reaches $0$, the label $l_n$ no longer propagates onward (see~\eqref{alpa}), which successfully eliminates the emergence of a major community.

Label hop attenuation can be rewritten into an equivalent form that allows altering $\delta$ during the course of the algorithm~\cite{LHLC09}. One keeps the label distance from the origin $d_n$ (initially set to $0$) that is updated after each propagation. Hence,
\begin{eqnarray}
d_n=\left(\min_{m\in\mathcal{N}^{c_n}_n}d_m\right)+1,
\end{eqnarray}
when the score $s_n$ is then
\begin{eqnarray}
s_n=1-\delta d_n.
\end{eqnarray}

Further analysis in~\cite{LHLC09} has revealed that label hop attenuation has to be coupled with \textit{node preference} $f_m$ (i.e. node propagation strength), in order for the algorithm to improve on the basic label propagation. Thus, the label propagation updating rule in~\eqref{lpa} is transformed into
\begin{eqnarray}
c_n=\argmax_l\sum_{m\in\mathcal{N}^l_n}f_m^\alpha s_mw_{nm},
\label{eq_alpa}
\end{eqnarray}
where $\alpha$ is a parameter of the algorithm. Leung~et~al.~\cite{LHLC09} have experimented with node preference equal to the degree of the node (i.e. $f_m=deg_m$ and $\alpha=0.1$), however, no general analysis was conducted.

The updating rule of label propagation (\eqref{lpa}), or its refinements (\eqref{alpa}), might prevent the algorithm from converging~\cite{RAK07}. Imagine a \textit{bipartite network} with two sets of nodes, i.e. red and blue nodes. Let, at some iteration of the algorithm, all red nodes share label $l_r$ and all blue nodes share label $l_b$. Due to the bipartite structure of the network, at the next iteration, all red, blue nodes will adopt label $l_b$, $l_r$ respectively. Furthermore, after the next iteration, all nodes will recover their original labels, failing the algorithm to converge.

The problem can be avoided by using \textit{asynchronous} updating~\cite{RAK07}. Nodes are no longer updated all together, but sequentially, in a random order. Thus, when node's label is updated, (possibly) already updated labels of its neighbors are considered (in contrast to \textit{synchronous} updating that considers only labels from the previous iteration). All of the algorithms, presented in the following section, use such asynchronous updating of nodes.

\section{\label{sec_defoff}Defensive and offensive label propagation}
In this section we present different algorithms that employ two unique strategies of community formation, namely, \textit{defensive preservation} and \textit{offensive expansion} of communities. First, we briefly present a \textit{dynamic hop attenuation} technique in~\secref{hopatt}. Next, \secref{strats} introduces and formally discusses the two strategies and associated algorithms (denoted \textit{dDaLPA} and \textit{oDaLPA} respectively). Last, \secref{kcores} presents an advanced label propagation algorithm (denoted \textit{K-Cores}) that combines the two strategies in an iterative manner, thus retaining the advantages of both approaches.

\subsection{\label{sec_hopatt}Dynamic hop attenuation}
Label hop attenuation has proven to be a reliable technique for prevention of emergence of a major community (\secref{lpa}). Still, it is not immediately evident what should the value of attenuation ratio $\delta$ be. In~\cite{LHLC09} authors have obtained good results with values around $0.1$, however, only a limited set of networks was considered.

We propose a \textit{dynamic hop attenuation} technique based on the hypothesis\footnotemark[3] that hop attenuation should only be employed when a label, or a set of labels, rapidly occupies a large portion of the network (which could potentially result in a formation of a major community). Otherwise, the restriction should be (almost) completely relaxed to allow label propagation to reach the equilibrium unrestrained. The technique would thus retain the dynamics of label propagation, but still successfully prevent the emergence of a major community. 

\footnotetext[3]{Similar idea was already discussed in~\cite{LHLC09}.}

We employ the following hop attenuation strategy. After each iteration (i.e. sweep through all the nodes) $\delta$ is set to the proportion of nodes that changed their labels\footnotemark[4] (on the first two iterations $\delta$ is set to $0.5$ and $0.1$ respectively). In practice, this results in higher values of $\delta$ in the early iterations of the algorithm, which enables the occurrence of a larger number of (smaller) well defined communities, when in the later stages $\delta$ gradually converges to $0$, which refines the communities and preserves only those strongly depicted in the network topology. Moreover, empirical analysis on real-world networks shows that such dynamic strategy successfully eliminates the emergence of a major community (the exact results are omitted).

\footnotetext[4]{The proportion of nodes that change their labels on the first five iterations roughly follows the sequence $90\%$, $30\%$, $10\%$, $5\%$, $3\%$~\cite{SB00} (on networks of moderate size).}

Note that an additional constraint should be imposed to prevent extremely large values of $\delta$ in the late stages of the algorithm (which, due to the above discussion, indicates some spurious behavior). Thus, to ensure convergence, when $\delta$ is greater than $\delta_{max}$ (e.g. $\delta_{max}=0.5$), we set it to $0$.

\subsection{\label{sec_strats}Defensive preservation and offensive expansion of communities}
Leung~et~al.~\cite{LHLC09} have shown that applying node preference (\secref{lpa}), to alter propagation strength or spread from certain nodes, can greatly improve the performance of the basic label propagation. Nevertheless, our empirical analysis has revealed that different networks favor different strategies for node preference~\cite{SB00}. 

On small social networks, where high degree nodes reside in the core of each community (e.g. Zachary's karate club network~\cite{Zac77}), good performance can be obtained by using \textit{degree} or \textit{eigenvector centrality}~\cite{Fre77,Fre79} for node preference. However, on Girvan~and~Newman~\cite{GN02} benchmark networks with planted partition, where all nodes have equal degree (on average), the measures render useless and are outperformed by \textit{clustering coefficient}~\cite{WS98}. Furthermore, on Lancichinetti~et~al.~\cite{LFR08} benchmark networks superior performance is obtained by using inverted degree or inverted eigenvector centrality. Interestingly, the measures thus complement each node's degree, decreasing the propagation strength from high degree nodes (and vice-versa). In summary, the analysis has revealed that none of the considered measures is appropriate for general networks (all different kinds of networks).

We have observed that, during the course of the algorithm, applying node preference to the core of each current community (i.e. to its most central nodes) can significantly increase the performance on a wide range of real-world networks. Furthermore, the strategy results in a great ability of detecting communities, even when they are only weakly defined. On the other hand, applying node preference to the border of each current community (i.e. to its edge nodes) results in an extremely accurate detection, expanding communities that are strongly depicted in the network topology.

Based on above observations we propose two algorithms that convey two unique strategies of community formation. The algorithms estimate the core (and border) of each (current) community by means of the \textit{diffusion} over the network; and are denoted \textit{defensive} and \textit{offensive diffusion label propagation algorithm} ($dDaLPA$ and $oDaLPA$ respectively). Let $p_n\in (0,1)$ be a value for node $n\in N$ thus that nodes in the core of the community have higher values of $p_n$ than border nodes. The defensive algorithm $dDaLPA$ applies preference (i.e. propagation strength) to the core of each community, i.e. $f_n^\alpha=p_n$, and the updating rule in \eqref{alpa} rewrites to
\begin{eqnarray}
c_n=\argmax_l\sum_{m\in\mathcal{N}^l_n}p_ms_mw_{nm}.
\label{eq_ddalpa}
\end{eqnarray}
On the other hand, the offensive version $oDaLPA$ applies preference to the border of each community, i.e. $f_n^\alpha=1-p_n$, and the updating rule becomes
\begin{eqnarray}
c_n=\argmax_l\sum_{m\in\mathcal{N}^l_n}(1-p_m)s_mw_{nm}.
\label{eq_odalpa}
\end{eqnarray}
During the course of the algorithm, values $p_n$ are estimated using \textit{random walks} within each current community. Let $p_n$ be the probability that a random walker, utilized on the community labeled with $c_n$, visits node $n$ (due to simplicity, we assume that community features \textit{connectedness}). $p_n$ can then be computed as
\begin{eqnarray}
p_n=\sum_{m\in\mathcal{N}^{c_n}_n}p_m/deg_m^{c_n},
\label{eq_diff}
\end{eqnarray}
where $deg_m^{c_n}$ is the intra-community degree of node $m$ ($c_m=c_n$). Besides deriving an estimate of the core and border of each community, the rationale here is to formulate label propagation (i.e. diffusion) within each of the current communities. Thus, opposed to the algorithm in~\cite{LHLC09}, the main novelty is in considering (current) communities, found by the algorithm, to estimate the (current) state of the label propagation process and then to adequately alter the dynamics of the process.

\begin{figure}[t]
\centering
\includegraphics[width=1.00\textwidth]{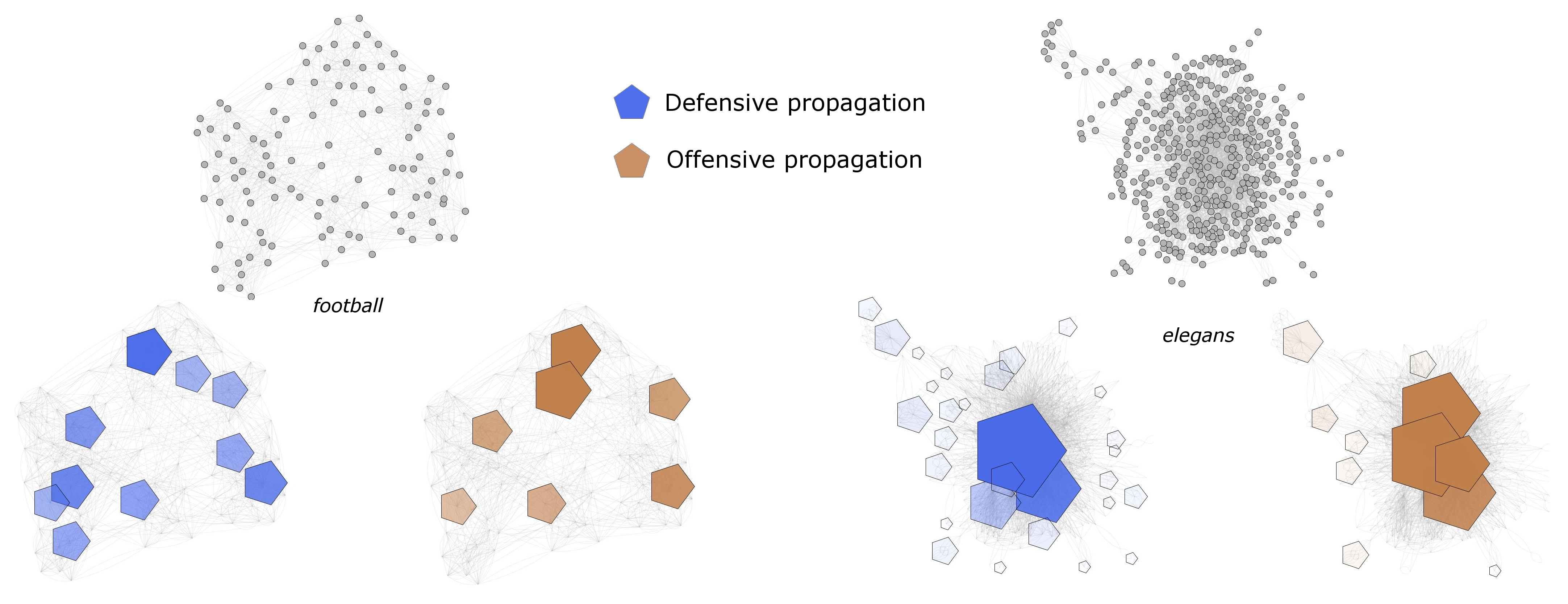}
\caption{\label{fig_defoff}Comparison of defensive and offensive label propagation on two real-world networks (see~\tblref{eval_desc}). Revealed communities are shown with pentagonal nodes, when the sizes (and colors) of nodes are proportional to the sizes of communities. Defensive propagation produces a larger set of communities that are (on average) considerably smaller than those revealed by the offensive propagation.}
\end{figure}

Defensive and offensive label propagation algorithms result in two unique strategies of community formation, namely, \textit{defensive preservation} and \textit{offensive expansion} of communities. The defensive algorithm quickly establishes a larger number of strong community cores (in the sense of \eqref{ddalpa}) and is able to defensibly preserve them during the course of the algorithm. This results in an immense ability of detecting communities, even when they are only weakly defined in the network topology. On the other hand, the offensive approach produces a much smaller set of communities (of various sizes). Laying the pressure on the edge of each community expands (i.e. enlarges) those that are strongly depicted in the network topology. This constitutes a more natural (offensive) struggle among communities and results in a great accuracy of the communities revealed.

Comparison of the approaches on two real-world networks is shown in~\figref{defoff}; and for pseudo-code of the algorithms see~\algref{dalpa}.

\begin{algorithm}
\algsetup{indent=1em}
\caption{\label{alg_dalpa}Defensive label propagation algorithm (\textit{dDaLPA}).}
\begin{algorithmic}[0]
\REQUIRE Undirected graph $G(N,E)$ with weights $W$
\ENSURE Communities $C$ (i.e. node labels)

\STATE $\delta\gets 0.5$
\FOR{$n\in N$} 
	\STATE $c_n\gets l_n$ \COMMENT{Unique label.}
	\STATE $p_n\gets 1/|N|$
	\STATE $d_n\gets 0$
\ENDFOR

\WHILE{\NOT \textit{converged}}
	\STATE \textit{shuffle}$(N)$
	\FOR{$n\in N$} 
		\STATE $c_n\gets \argmax_l\sum_{m\in\mathcal{N}^l_n}p_m(1-\delta d_m)w_{nm}$ \COMMENT{$1-p_m$ instead of $p_m$ for \textit{oDaLPA}.}
		\STATE $p_n\gets \sum_{m\in\mathcal{N}^{c_n}_n}p_m/deg^{c_n}_m$ \COMMENT{$deg_m$ instead of $deg^{c_n}_m$ for \textit{oDaLPA}.}
		\IF{$c_n$ \textit{has changed}}
			\STATE $d_n\gets (\min_{m\in\mathcal{N}^{c_n}_n}d_m)+1$
		\ENDIF
	\ENDFOR
	\STATE $\delta\gets$ \textit{proportion of labels changed} \COMMENT{$\delta\gets 0.1$ at the first iteration.}
	\IF{$\delta\geq\delta_{max}$} 
		\STATE $\delta\gets 0$ \COMMENT{Ensuring convergence ($\delta_{max}$ is fixed to $0.5$).}
	\ENDIF
\ENDWHILE

\RETURN $C$  \COMMENT{Returns (relabeled) communities that feature connectedness.}
\end{algorithmic}
\end{algorithm}

\subsection{\label{sec_kcores}Combining defensive and offensive propagation}
Defensive and offensive label propagation (\secref{strats}) convey two unique strategies of community formation. An obvious improvement would be to combine the strategies, thus retaining the strong detection ability of the defensive approach and high accuracy of the offensive strategy. However, simply using the algorithms one after another does not attain the desired properties -- any label propagation algorithm, being run until convergence, finds a local optimum (i.e. local equilibrium) that is hard to escape from. 

Raghavan~et~al.~\cite{RAK07} have already discussed the idea (however, in different context) that label propagation could be improved, if one had \textit{a priori} knowledge about community cores. Core nodes could then be labeled with the same label, leaving all the other nodes labeled with an unique label. During the course of the algorithm, the (uniquely labeled) nodes would tend to adopt the label of their nearest \textit{attractor} (i.e. community core) and thus join its community. This would improve the algorithm's stability~\cite{RAK07} and also the accuracy of the identified communities (\secref{eval}).

\begin{figure}[t]
\centering
\includegraphics[width=0.85\textwidth]{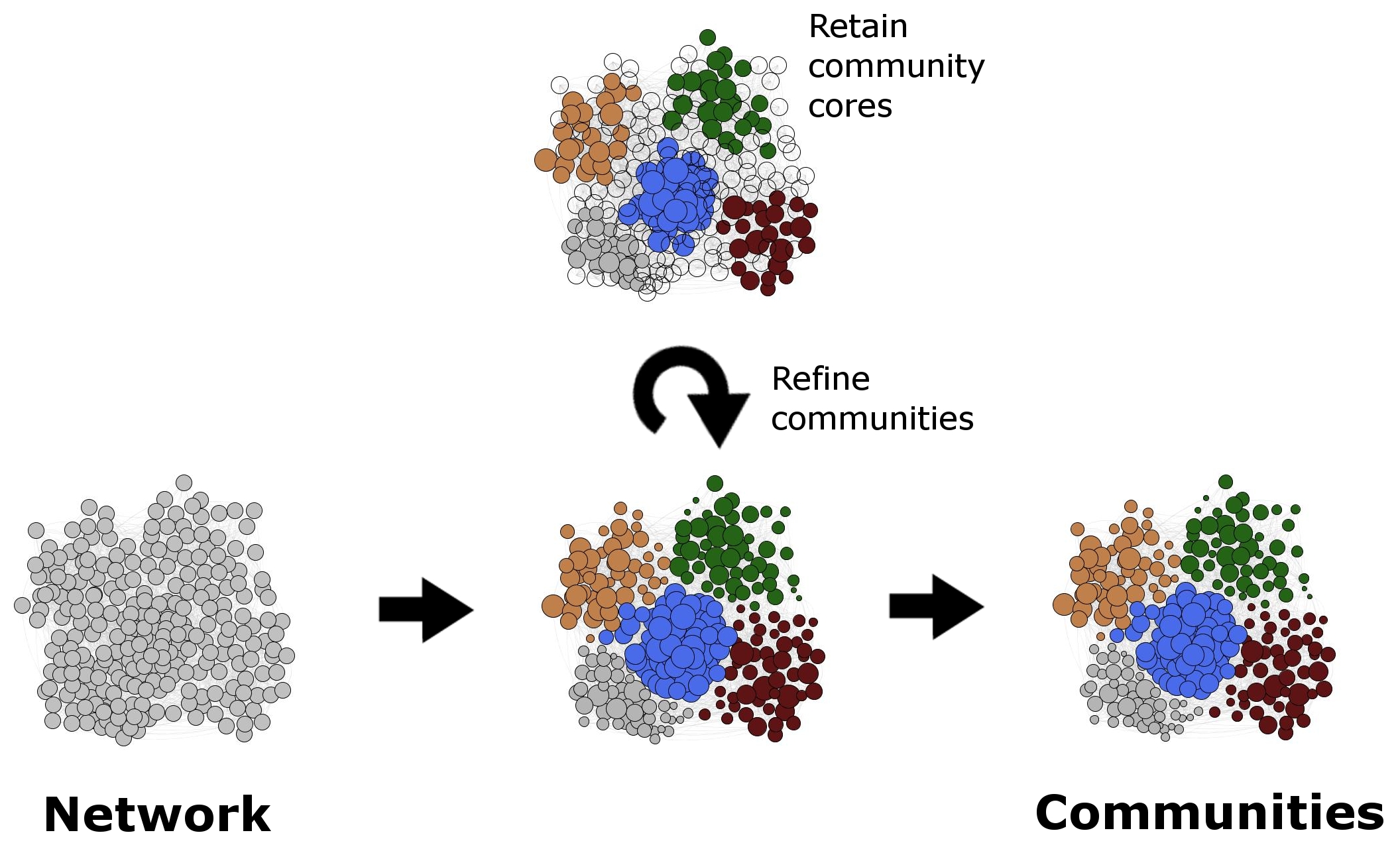}
\caption{\label{fig_kcores}Schematic representation of \textit{K-Cores} algorithm.}
\end{figure}

The defensive and offensive label propagation algorithms are thus combined in the following manner (\figref{kcores}). First, the defensive strategy is applied, to produce initial estimates of the communities and to accurately detect their cores. All border nodes of each community are then relabeled (labeled with unique labels), so that (approximately) one half of the nodes retain their original label. Next, the offensive strategy is applied, which refines the community cores and accurately detects also their borders. Relabeling and offensive refinement are then repeated until the number of communities decreases. Such combined strategy preserves advantages of both, defensive and offensive, label propagation (\secref{eval}) and is denoted \textit{K-Cores}\footnotemark[5] algorithm (due to its resemblance to a well-known \textit{K-Means} algorithm~\cite{Mac67}). 

\footnotetext[5]{The term should not be confused with \textit{$k$-core}~\cite{Sei83} that denotes the maximal subgraph in which each node has degree at least $k$.}

The core (and border) of each community is again estimated by means of (diffusion) values $p_n$ (\secref{strats}). Thus, within the algorithm, the node $n$ is relabeled due to the following rule,
\begin{subnumcases}{c_n=}
c_n & for $p_n>m_{c_n}$\\
l_n & for $p_n\leq m_{c_n}$,
\label{eq_relab}
\end{subnumcases}
where $m_{c_n}$ is the \textit{median} of values $p_n$, for nodes in the community $c_n$, and $l_n$ is an unique label. Hence, the core nodes retain their original labels, when all border nodes are relabeled.

Schematic representation of the algorithm is depicted in \figref{kcores}; and for the pseudo-code of the algorithm see~\algref{kcores}. For a further discussion on all presented algorithms see~\cite{SB00}.

\begin{algorithm}
\algsetup{indent=1em}
\caption{\label{alg_kcores}\textit{K-Cores} algorithm.}
\begin{algorithmic}[0]
\REQUIRE Undirected graph $G(N,E)$ with weights $W$
\ENSURE Communities $C$ (i.e. node labels)

\STATE $C\gets$ \textit{dDaLPA}$(G,W)$ \COMMENT{Defensive label propagation.}
\WHILE{$|C|$ \textit{decreases}} 
	\FOR{$c\in C$} 
		\STATE $m_c\gets$\textit{median}$(\{p_n|\mbox{ }n\in N\wedge c_n=c\})$ \COMMENT{Retain community cores.}
		\FOR{$n\in N$ \AND $c_n=c$} 
			\IF{$p_n\leq m_c$}
				\STATE $c_n\gets l_n$ \COMMENT{Unique label.}
				\STATE $p_n\gets 1/|N|$
			\ENDIF
			\STATE $d_n\gets 0$
		\ENDFOR
	\ENDFOR
	\STATE $C\gets$ \textit{oDaLPA}$(G,W)$ \COMMENT{Offensive label propagation.}
\ENDWHILE

\RETURN $C$ \COMMENT{Returns best communities found.}
\end{algorithmic}
\end{algorithm}

\section{\label{sec_eval}Empirical evaluation and discussion}
In this section we present and discuss results of the empirical evaluation of the proposed algorithms. \Secref{eval_benchmarks} gives results of the analysis on benchmark networks with planted partition, when the results on real-world network are reported in~\secref{eval_rws}. For the use with larger networks, we also briefly present and empirically compare two manners of hierarchical community detection in~\secref{eval_large}.

The results are assessed using two measures of community structure, namely, \textit{Normalized Mutual Information} \textit{NMI}~\cite{DDDA05} and \textit{modularity} $Q$~\cite{NG04}. The latter measures the relative significance of the communities due to a selected \textit{null model}. Let $A_{nm}$ denote the number of edges incident to nodes $n, m\in N$ and let $P_{nm}$ be the expected number of incident edges in the null model. The modularity then reads
\begin{eqnarray}
Q & = & \frac{1}{2|E|}\sum_{n,m\in N}\left(A_{nm}-P_{nm}\right)\delta(c_n,c_m),
\end{eqnarray}
where $c_n$ is the identified community (label) for node $n\in N$ and $\delta$ is the Kronecker delta. The modularity thus measures the fraction of the difference between the number intra-community edges and the expected number of edges in the null model ($Q\in [-1,1]$). Commonly a random graph with the same degree distribution as the original is selected for the null model. Hence, $P_{nm}=\frac{deg_ndeg_m}{2|E|}$. 

Furthermore, the analysis on networks with planted partition is conducted using \textit{Normalized Mutual Information} \textit{NMI}~\cite{DDDA05}. Let $\mathcal{C}$ be the partition (i.e. communities) extracted by some algorithm and let $\mathcal{P}$ be the planted partition of the network (corresponding random variables are $C$ and $P$ respectively). The \textit{NMI} of $\mathcal{C}$ and $\mathcal{P}$ is then
\begin{eqnarray}
\mbox{\textit{NMI}} & = & \frac{2I(C,P)}{H(C)+H(P)},
\end{eqnarray}
where $I(C,P)$ is the \textit{mutual information} of the partitions, $I(C,P)=H(C)-H(C|P)$, and $H(C)$, $H(P)$ and $H(C|P)$ are standard and conditional entropies. \textit{NMI} of identical partitions equals $1$, and is $0$ for independent partitions.

\subsection{\label{sec_eval_benchmarks}Networks with planted partition}
\begin{figure}[p]
\centering
\includegraphics[width=1.00\textwidth]{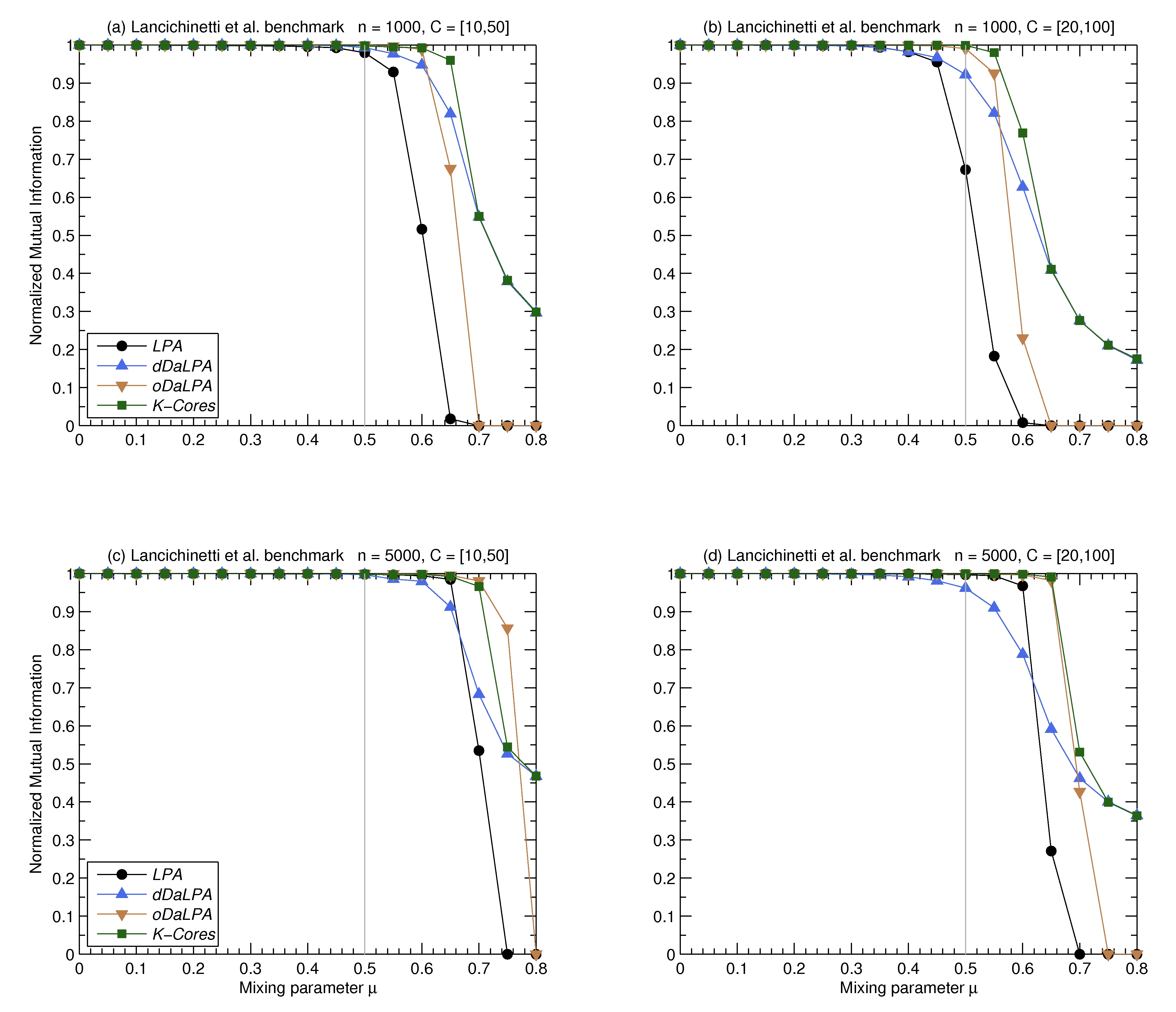}
\caption{\label{fig_benchmarks}Comparison of the proposed algorithms on Lancichinetti~et~al.~\cite{LFR08} benchmarks networks with planted partition. The network sizes equal $1000$ and $5000$ nodes respectively; and communities comprise of up to $50$ and $100$ nodes respectively. The results were averaged over $100$ realizations of the benchmarks networks.}
\end{figure}

The proposed algorithms were first analyzed on four different types of Lancichinetti~et al.~\cite{LFR08} benchmark networks with planted partition. The results are shown in~\figref{benchmarks}.

\begin{figure}[p]
\centering
\includegraphics[width=0.45\textwidth]{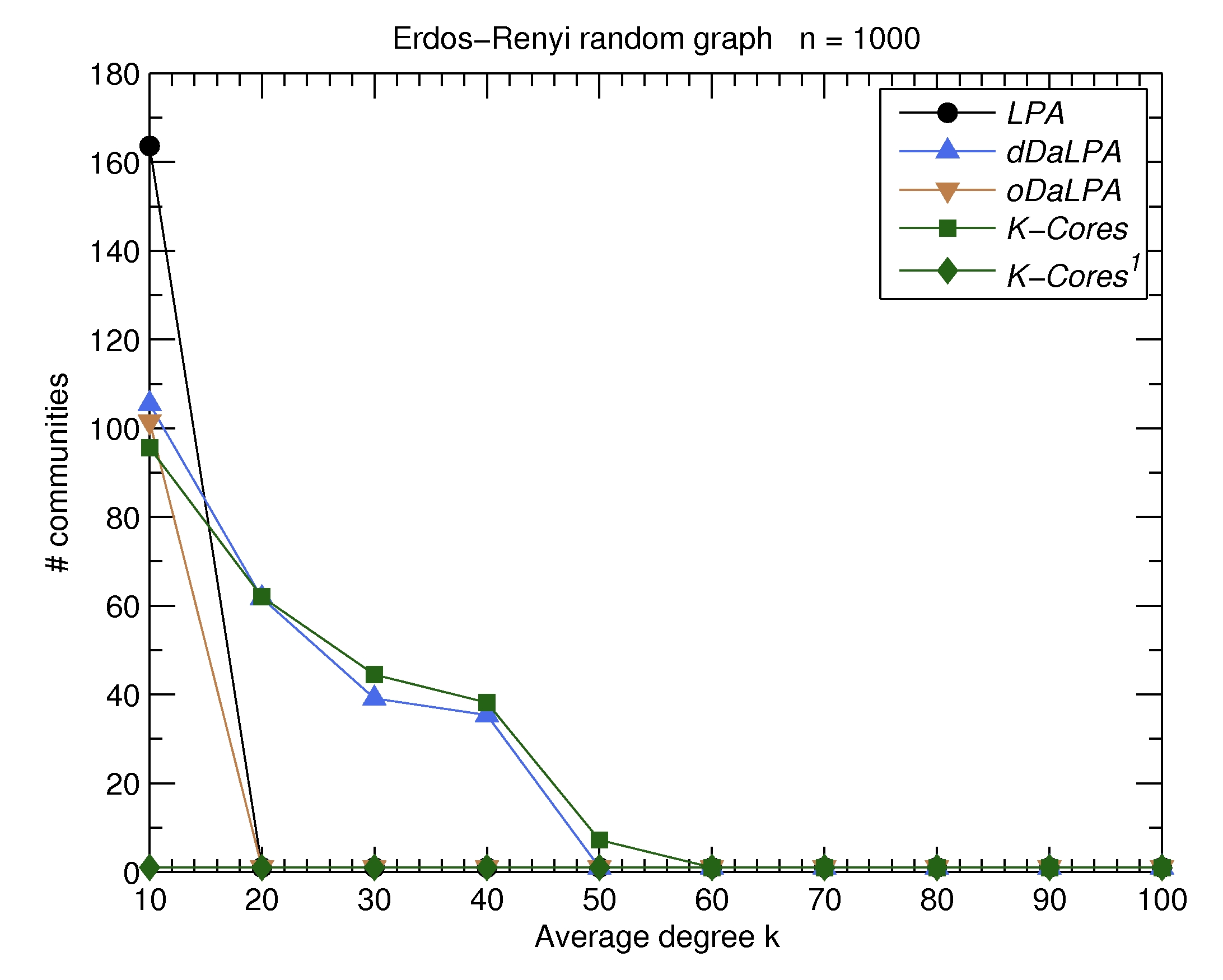}
\caption{\label{fig_random}Comparison of the proposed algorithms on a random graph \`{a}~la~Erd\"{o}s-R\'{e}nyi~\cite{ER59} with $1000$ nodes (the results were averaged over $10$ runs).}
\end{figure}

The analysis clearly depicts the difference between defensive and offensive label propagation. The offensive approach (\textit{oDaLPA}) performs considerably better than the basic label propagation (\textit{LPA}) and can still accurately detect communities, when \textit{LPA} already fails. On the other hand, the defensive propagation (\textit{dDaLPA}) does not detect communities as accurately as the offensive approach, and \textit{LPA} on larger networks, but still reveals communities, even when they are only weakly defined. Furthermore, \textit{K-Cores} algorithm outperforms all other approaches in all but one case. Note that the algorithm retains the advantages of both defensive and offensive approach, still, the performance does not simply equal to the \textit{upper-hull} of those for \textit{dDaLPA} and \textit{oDaLPA}.

The algorithms were also applied to a random graph \`{a}~la~Erd\"{o}s-R\'{e}nyi~\cite{ER59} that (presumably) has no community structure (\figref{random}). However, the defensive label propagation still reports communities, when the average degree is small enough. Nevertheless, further analysis reveals that defensive label propagation is still a preferred approach on a wide range of real-world networks (\secref{eval_rws}). 

As \textit{K-Cores} algorithm is initialized using the defensive propagation, and best communities are reported at the end, the performance for \textit{K-Cores} on a random graph is similar to that for \textit{dDaLPA}. However, if we discard the initial communities obtained by~\textit{dDaLPA} (i.e. \textit{K-Cores$^1$} algorithm), the results correspond to those for \textit{LPA} and \textit{oDaLPA} that reveal no community structure.

\subsection{\label{sec_eval_rws}Real-world networks}
The algorithms were further analyzed on over $20$ real-world networks of moderate size (\tblref{eval_rws}). Due to a large number of networks considered, the detailed description is omitted (see~\tblref{eval_desc}). However, the set includes different communication, social, biological, web, (author) collaboration, Internet and other networks. Due to simplicity, all networks are considered as unweighted and undirected, i.e. all weighted or directed edges are treated as simple undirected edges (same holds for networks in~\secref{eval_large}).

\begin{table}[p]
\centering
\caption{\label{tbl_eval_desc}Networks used for the analysis of community detection algorithms.}
\begin{tabular}{ccc}
\hline\noalign{\smallskip}
Network & Description & Reference \\
\noalign{\smallskip}
\hline
\noalign{\smallskip}
\textit{uni} & Emails within an university. & \cite{GDDGA03} \\
\textit{enron} & Emails within \textit{Enron}. & \cite{LKF05} \\
\noalign{\smallskip}
\hline
\noalign{\smallskip}
\textit{football} & American college football league. & \cite{GN02} \\
\textit{jazz} & Network of jazz musicians. & \cite{GD03} \\
\textit{wiki} & Voting network of \textit{Wikipedia}. & \cite{LHK10} \\
\textit{epinions} & \textit{Epinions} web of trust. & \cite{RAD03} \\
\noalign{\smallskip}
\hline
\noalign{\smallskip}
\textit{yeast} & Yeast protein interactions. & \cite{JMBO01} \\
\noalign{\smallskip}
\hline
\noalign{\smallskip}
\textit{elegans} & Nematode \textit{Caenorhabditis elegans}. & \cite{JTAOB00} \\
\noalign{\smallskip}
\hline
\noalign{\smallskip}
\textit{gnutella} & \textit{Gnutella peer-to-peer} network. & \cite{LKF07} \\
\noalign{\smallskip}
\hline
\noalign{\smallskip}
\textit{blogs} & Weblogs on U.S. politics. & \cite{AG05} \\
\noalign{\smallskip}
\hline
\noalign{\smallskip}
\textit{genrelat} & \textit{General Relativity} archive 2003. & \cite{LKF07}  \\
\textit{codmat$^3$} & \textit{Condensed Matter} archive 2003. & \cite{New01} \\
\textit{codmat$^5$} & \textit{Condensed Matter} archive 2005. & \cite{New01} \\
\textit{hep} & \textit{High Energy Physics} archive 2003. & \cite{LKF07} \\
\textit{astro} & \textit{Astro Physics} archive 2003. & \cite{LKF07} \\
\noalign{\smallskip}
\hline
\noalign{\smallskip}
\textit{engine} & \textit{Google App Engine} library. \\
\textit{jung} & \textit{JUNG} graph and network library. \\
\textit{javax} & \textit{Java 6} \textit{javax} namespace. \\
\noalign{\smallskip}
\hline
\noalign{\smallskip}
\textit{power} & Western U.S. power grid. & \cite{WS98} \\
\noalign{\smallskip}
\hline
\noalign{\smallskip}
\textit{oregon$^3$} & Aut. syst. of Internet 2003 (\textit{Oregon}). & \cite{LKF05} \\
\textit{oregon$^6$} & Aut. syst. of Internet 2006 (\textit{Oregon}). & \cite{New06b} \\
\textit{nec} & \textit{nec} web overlay map. & \cite{HJJMMMM03} \\
\noalign{\smallskip}
\hline
\noalign{\smallskip}
\textit{amazon} & \textit{Amazon} co-purchasing network. & \cite{LAH07} \\
\noalign{\smallskip}
\textit{ndedu} & Web graph of \textit{nd.edu} domain. & \cite{AJB99} \\
\noalign{\smallskip}
\textit{road} & Roads in Pennsylvania. & \cite{LLDM08} \\
\noalign{\smallskip}
\textit{google} & Web graph of \textit{Google}. & \cite{LLDM08} \\
\noalign{\smallskip}
\textit{skitter} & Aut. syst. of Internet 2005 (\textit{Skitter}). & \cite{LKF05} \\
\noalign{\smallskip}
\textit{movie} & Movie actors collaborations. & \cite{BA99} \\
\noalign{\smallskip}
\textit{nber} & \textit{NBER} patents citations. & \cite{HJT01} \\
\noalign{\smallskip}
\textit{live} & \textit{Live Journal} friendships. & \cite{LLDM08} \\
\noalign{\smallskip}
\textit{webbase} & Web graph from \textit{WebBase}. & \cite{Aut10} \\
\hline
\end{tabular}
\end{table}

\begin{figure}[h]
\centering
\includegraphics[width=0.95\textwidth]{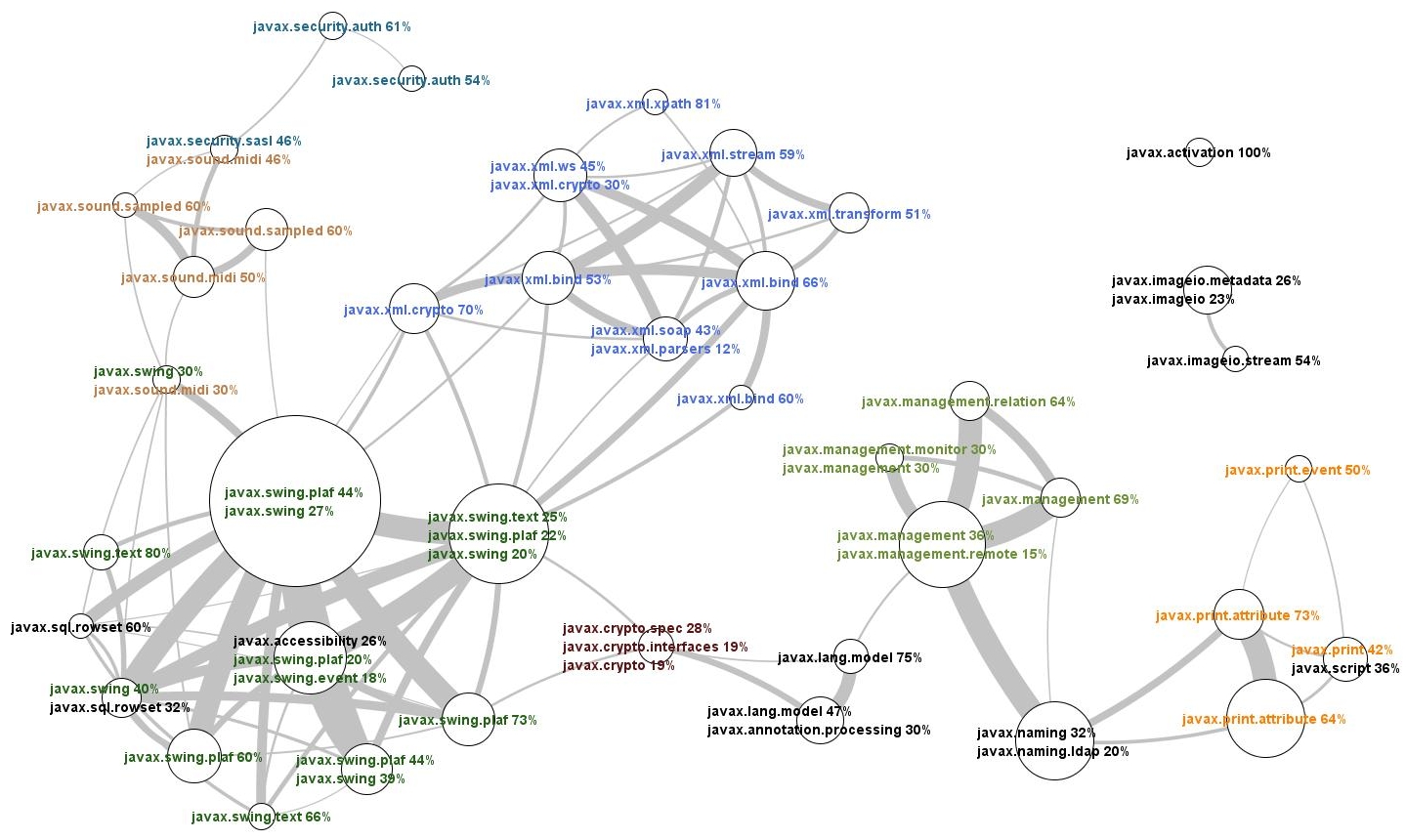}
\caption{\label{fig_javax}Communities revealed in \textit{javax} software network by applying \textit{K-Cores} algorithm. The sizes of nodes correspond to the sizes of communities; and the widths of the edges correspond to the number of inter-community edges (due to clarity, weakly represented nodes and edges were discarded). Text shows the distribution of \textit{javax} packages within the communities, where all weakly represented packages were omitted.}
\end{figure}

We also introduce a new type of networks denoted \textit{software networks} (sort of \textit{component dependency networks}~\cite{WKD07}). Here nodes represent a set of classes of some software system, written in an object-oriented programing language, and edges represent relations among them. Two classes $A$ and $B$ are defined as related when $B$ extends or implements $A$, when $B$ contains a field of type $A$ or when $A$, $B$ contains a method that returns, requires an object of type $B$, $A$ respectively. The hypothesis here is that network communities would correspond to software packages, which could result in numerous applications in software engineering domain (\figref{javax}). In this article we consider the ground case, where networks are represented with simple undirected and unweighted graphs\footnotemark[6].

\footnotetext[6]{The networks were obtained by parsing the documentation of the corresponding software. Thus, due to various reasons, some false relations might have been introduced.}

\begin{table}
\centering
\caption{\label{tbl_eval_rws} Mean modularities $Q$ for different label propagation algorithms (averaged over $100$ to $100000$ runs). The highest values of $Q$ are shown with solid font; and underlined values correspond to the highest values among only \textit{dDaLPA} and \textit{oDaLPA}.}
\begin{tabular}{ccrrccccc}
\hline\noalign{\smallskip}
Type & Network & Nodes & Edges & $ $ $ $ & $ $ $ $ \textit{LPA} $ $ $ $ & \textit{dDaLPA} & \textit{oDaLPA} & \textit{K-Cores}\\
\noalign{\smallskip}
\hline
\noalign{\smallskip}
\multirow{2}{*}{Communication} & \textit{uni} & 1133 & 5451&  & 0.364 & \underline{0.481} & 0.389 & \textbf{0.518} \\
 & \textit{enron} & 36692 & 367662&  & 0.355 & \underline{0.514} & 0.380 & \textbf{0.516} \\
\noalign{\smallskip}
\hline
\noalign{\smallskip}
\multirow{4}{*}{Social} & \textit{football} & 115 & 616 & & 0.592 & 0.593 & \underline{0.595} & \textbf{0.600} \\
 & \textit{jazz} & 198 & 2742 & & 0.346& \underline{\textbf{0.418}} & 0.377 & \textbf{0.418} \\
 & \textit{wiki} & 7115 & 103689 & & 0.056 & \underline{0.195} & 0.046 & \textbf{0.202} \\
 & \textit{epinions} & 75879 & 508837 & & 0.106 & \underline{0.288} & 0.111 & \textbf{0.291} \\
\noalign{\smallskip}
\hline
\noalign{\smallskip}
\multirow{1}{*}{Protein} & \textit{yeast} & 2114 & 4480 & &  0.665 & \underline{0.733} & 0.720 & \textbf{0.793} \\
\noalign{\smallskip}
\hline
\noalign{\smallskip}
\multirow{1}{*}{Metabolic} & \textit{elegans} & 453 & 2025 & &  0.122 & \underline{0.172} & 0.131 & \textbf{0.173} \\
\noalign{\smallskip}
\hline
\noalign{\smallskip}
\multirow{1}{*}{Peer-to-peer} & \textit{gnutella} & 62586 & 147892 & &  0.338 & \underline{0.412} & 0.387 & \textbf{0.447} \\
\noalign{\smallskip}
\hline
\noalign{\smallskip}
\multirow{1}{*}{Web} & \textit{blogs} & 1490 & 16718 & &  0.400 & 0.424 & 0.424 & \textbf{0.426} \\
\noalign{\smallskip}
\hline
\noalign{\smallskip}
\multirow{5}{*}{Collaboration} & \textit{genrelat} & 5242 & 28980 & & 0.737 & 0.769 & \underline{0.779} & \textbf{0.820} \\
 & \textit{codmat$^3$} & 27519 & 116181 & & 0.596 & 0.611 & \underline{0.627} & \textbf{0.687} \\
 & \textit{codmat$^5$} & 36458 & 171736 & & 0.548 & 0.575 & \underline{0.590} & \textbf{0.648} \\
 & \textit{hep} & 12008 & 237010 & & 0.484 & \underline{\textbf{0.585}} & 0.518 & \textbf{0.585} \\
 & \textit{astro} & 18772 & 396160 & & 0.326 & \underline{\textbf{0.538}} & 0.337 & \textbf{0.538} \\
\noalign{\smallskip}
\hline
\noalign{\smallskip}
\multirow{3}{*}{Software} & \textit{engine} & 139 & 243 & &  0.689 & 0.724 & \underline{0.726} & \textbf{0.747} \\
 & \textit{jung} & 436 & 1303 & &  0.611 & 0.587 & \underline{0.623} & \textbf{0.631} \\
 & \textit{javax} & 2089 & 7934 & &  0.723 & 0.687 & \underline{0.725} & \textbf{0.768} \\
\noalign{\smallskip}
\hline
\noalign{\smallskip}
\multirow{1}{*}{Power} & \textit{power} & 4941 & 6594 & & 0.595 & 0.690 & \underline{0.698} & \textbf{0.820} \\
\noalign{\smallskip}
\hline
\noalign{\smallskip}
\multirow{3}{*}{Internet}  & \textit{oregon$^3$} & 767 & 3591 & & 0.302 & 0.210 & \underline{\textbf{0.354}} & 0.210 \\
 & \textit{oregon$^6$} & 22963 & 48436 & & 0.498 & 0.347 & \underline{\textbf{0.541}} & 0.347 \\
 & \textit{nec} & 75885 & 357317 & & 0.683 & 0.628 & \underline{0.688} & \textbf{0.736} \\
\hline
\end{tabular}
\end{table}

Comparison of the algorithms on real-world networks (\tblref{eval_rws}) firstly confirms the adequacy of the \textit{K-Cores} algorithm that obtains highest modularity on all but two Internet networks. Note that both \textit{dDaLPA} and \textit{oDaLPA} also outperform the basic \textit{LPA} in most cases. Moreover, the analysis clearly separates different types of networks due to the preferred strategy of community formation. For instance, social and communication networks clearly favor defensive preservation of communities, due to high density of such networks (and rather weakly defined communities). On the other hand, sparse software or Internet networks, with longer paths among nodes, obviously prefer the offensive expansion of communities. The middle case is represented by the considered collaboration networks. On smaller networks that are relatively sparse (\textit{genrelat}, \textit{codmat}$^3$ and \textit{codmat}$^5$ network) the offensive approach prevails. However, on larger networks (\textit{hep} and \textit{astro} network), with significantly higher degrees then the former, the defensive algorithm is superior. In summary, denser networks (with higher average degrees) prefer the defensive preservation, whereas sparser networks (with lower average degrees) favor the offensive expansion of communities.

\subsection{\label{sec_eval_large}Analyzing large networks}
Last, we also briefly present and empirically evaluate two different manners of \textit{hierarchical} community investigation (prominent for the use with larger networks). Besides \textit{LPA} and \textit{K-Cores} algorithm, we consider the following approaches.

\textit{Basic diffusion and propagation algorithm} (\textit{DPA})~\cite{SB00} is an optimized version of \textit{K-Cores} that scales significantly better then the basic algorithm. Furthermore, \textit{hierarchical diffusion and propagation algorithm} (\textit{DPA}$^+$) represents a simple hierarchical detection, where the algorithm is recursively applied to the previously constructed \textit{community network}\footnotemark[7] (the algorithm employs defensive label propagation, and \textit{DPA} on the last step). Moreover, (general) \textit{diffusion and propagation algorithm} (\textit{DPA}$^*$)~\cite{SB00} represents a hierarchical \textit{core extraction} technique, where the algorithm recursively extracts the \textit{core}~\cite{LLDM08} of the network and identifies \textit{whisker} communities. For a further discussion on the algorithms see~\cite{SB00}.

\footnotetext[7]{A network whose nodes represent communities and edges represent edges between nodes in the original network.}

\begin{table}
\caption{\label{tbl_eval_large}Peak modularities $Q$ and average number of iterations for different label propagation algorithms (obtained over $1$ to $10$ runs). Solid values correspond to the largest values of $Q$, where missing values could not be obtained due to limited time resources.}
\centering
\begin{tabular}{crrccccccc}
\hline\noalign{\smallskip}
Network & Nodes & Edges & $ $ $ $ & \textit{LPA} & \textit{K-Cores} & \textit{DPA} & \textit{DPA$^+$} & \textit{DPA$^*$}\\
\noalign{\smallskip}
\hline
\noalign{\smallskip}
\textit{amazon} & 0.3M & 1.2M & &  0.681{\scriptsize/15} & 0.783{\scriptsize/273} & 0.700{\scriptsize/34} & \textbf{0.883{\scriptsize/65}} & 0.856{\scriptsize/78} \\
\noalign{\smallskip}
\textit{ndedu} & 0.3M & 1.5M & & 0.838{\scriptsize/53}  & 0.891{\scriptsize/471}  & 0.860{\scriptsize/50}  & 0.897{\scriptsize/37} & \textbf{0.901{\scriptsize/58}}  \\
\noalign{\smallskip}
\textit{road} & 1.1M & 3.1M & &  0.552{\scriptsize/10} & 0.847{\scriptsize/895} & 0.626{\scriptsize/82} & \textbf{0.985{\scriptsize/136}} & 0.883{\scriptsize/142} \\
\noalign{\smallskip}
\textit{google} & 0.9M & 4.3M & &  0.801{\scriptsize/15} & 0.889{\scriptsize/444} & 0.820{\scriptsize/59} & 0.962{\scriptsize/45} & \textbf{0.967{\scriptsize/48}} \\
\noalign{\smallskip}
\textit{skitter} & 1.7M & 11.1M & &  0.746{\scriptsize/25} & - & 0.755{\scriptsize/126} & 0.680{\scriptsize/52} & \textbf{0.801{\scriptsize/76}} \\
\noalign{\smallskip}
\textit{movie} & 0.4M & 15.0M & &  0.524{\scriptsize/21} & - & 0.533{\scriptsize/147} & 0.474{\scriptsize/39} & \textbf{0.606{\scriptsize/71}} \\
\noalign{\smallskip}
\textit{nber} & 3.8M & 16.5M & &  0.576{\scriptsize/109} & - & 0.582{\scriptsize/336} & 0.707{\scriptsize/112} & \textbf{0.739{\scriptsize/308}} \\
\noalign{\smallskip}
\textit{live} & 4.8M & 69.0M & & 0.673{\scriptsize/100} & - & 0.548{\scriptsize/206} & 0.683{\scriptsize/73} & \textbf{0.688{\scriptsize/125}} \\
\noalign{\smallskip}
\textit{webbase} & 14.5M & 101.0M & &  0.894{\scriptsize/38} & - & 0.923{\scriptsize/114} & 0.942{\scriptsize/43} & \textbf{0.954{\scriptsize/39}} \\
\hline
\end{tabular}
\end{table}

Algorithms were applied to a set of large real-world networks (\tblref{eval_large}) of various types (see~\tblref{eval_desc}), when the analysis on (even) larger networks was limited due to limited memory resources (i.e. $4$ GB of memory). On average, all of the considered algorithms again perform better then the basic label propagation (\textit{LPA}). Furthermore, the hierarchical algorithms (\textit{DPA}$^+$ and \textit{DPA}$^*$) obtain the highest values of modularity on all of the networks considered, whereas the core extraction technique (\textit{DPA}$^*$) seems more prominent.

The average number of iterations made by the algorithms\footnotemark[8] (\tblref{eval_large}) shows that ``theoretical'' approach \textit{K-Cores} does not scale to larger networks, where the optimized version \textit{DPA} is preferred. Note also that hierarchical detection even decreases the total number of iterations on larger networks, which gives promising grounds for future analysis of large complex networks.

\footnotetext[8]{As the employed implementation was optimized for memory, not time, resources, we report the number of iterations instead of the exact running times.}

For a further empirical evaluation and comparison with other label propagation algorithms reported in the literature see~\cite{SB00}.

\section{\label{sec_conc}Conclusion}
In the article we present different label propagation algorithms that employ two unique strategies of community formation, namely, defensive preservation and offensive expansion of communities. The strategies are combined in an advanced label propagation algorithm that retains the advantages of both approaches. Furthermore, we also show how the algorithm can be extended to larger networks using (hierarchical) core extraction. Nevertheless, the main contribution of this work is in showing that different types of networks (with different topological properties) favor different strategies of community formation.

Future work will focus mainly on further analyses of defensive and offensive label propagation, in order to develop an enhanced algorithm that would \textit{decide} between defensive and offensive strategy (an intermediate approaches) during the course of the algorithm. This could result in higher accuracy of the revealed communities and also in better scalability of the algorithm.


%
%

\bibliographystyle{splncs03}

\end{document}